\documentclass[12pt]{article}
\usepackage{graphics,cite,amssymb,epsfig,float,psfrag}
\usepackage[usenames,dvips]{color}
\usepackage{rotating}

\begin{document}

\newcommand{\lsim}{\raisebox{-0.13cm}{~\shortstack{$<$ \\[-0.07cm] $\sim$}}~}
\newcommand{\gsim}{\raisebox{-0.13cm}{~\shortstack{$>$ \\[-0.07cm] $\sim$}}~}

\baselineskip=14.8pt

\rightline{CERN-PH-TH/2012--186}
\rightline{DESY 12-113}
\rightline{LPT--Orsay 12/64}
\rightline{LPN 12-073}
\rightline{SFB/CPP-12-41}

\vspace{0.7cm}

\begin{center}

{\large\bf The top quark and Higgs boson masses and}  \\[1ex]
{\large\bf  the stability of the electroweak vacuum}

\vspace{.8cm}
  {\sc
    S.~Alekhin$^{\, a,}$\footnote{{\bf e-mail}: sergey.alekhin@ihep.ru},
    A.~Djouadi$^{\, b,c,}$\footnote{{\bf e-mail}: abdelhak.djouadi@th.u-psud.fr}
    and S.~Moch$^{\, d,}$\footnote{{\bf e-mail}: sven-olaf.moch@desy.de} \\
  }
  \vspace{0.6cm}
  {\small
    $^a$Institute for High Energy Physics, 
    142281 Protvino, Moscow region, Russia\\
    \vspace{0.2cm}
    $^b$Laboratoire de Physique Th\'eorique, CNRS and Universit\'e Paris-Sud, 
    F-91405 Orsay, France\\
    \vspace{0.2cm}
    $^c$ Theory Unit, Department of Physics, CERN, 
    CH-1211 Geneva 23, Switzerland\\
    \vspace{0.2cm}
    $^d$Deutsches Elektronensynchrotron DESY,
    Platanenallee 6, D-15738 Zeuthen, Germany \\
  }
\end{center}

\vspace{1.cm}

\begin{abstract} 

\noindent   The ATLAS and CMS experiments observed a particle at the LHC with a
mass  $\approx 126$ GeV, which is compatible with the Higgs boson of the Standard
Model.  A crucial question is, if for such a Higgs mass value, one could
extrapolate the model up to high scales while keeping the minimum of the scalar
potential that breaks the electroweak symmetry stable.   Vacuum stability
requires indeed the Higgs boson mass to be   $M_H\! \gsim \!  129 \pm 1$ GeV, but the 
precise value depends critically on the input top quark pole mass which is
usually taken to be the one measured at the Tevatron, $m_t^{\rm exp}=173.2 \pm
0.9$ GeV. However,  for an unambiguous and theoretically well-defined
determination of the top quark mass  one should rather use the total cross
section for top quark pair production at hadron colliders. Confronting  the
latest predictions of the  inclusive  $p \bar p \! \to \! t\bar t +X$ cross
section up to next-to-next-to-leading order in QCD to the experimental
measurement at the Tevatron, we determine the running mass in the $\overline{\rm
MS}$-scheme to be $m_t^{ \overline{\rm MS}}(m_t) = 163.3 \pm  2.7$~GeV which
gives a top quark pole mass of $m_t^{\rm pole }= 173.3 \pm  2.8$~GeV. This leads
to the vacuum stability constraint $M_H \geq 129.8 \pm 5.6$ GeV  to which a
$\approx 126$ GeV Higgs boson complies as the uncertainty is large.   A very
precise assessment of the stability of the electroweak vacuum  can only be made
at a future high-energy electron-positron collider,  where the top quark pole
mass could be determined with a few hundred MeV accuracy.   

\end{abstract}

\newpage
\setcounter{footnote}{0}

The recent results on Higgs boson searches delivered by the ATLAS and CMS
collaborations~\cite{LHC} at the Large Hadron Collider (LHC)  show that  there
is now an established signal (at almost five standard deviations for each
experiment) corresponding to a particle with a mass $\approx \! 126$ GeV and
with the properties expected for the  Standard Model (SM) Higgs
boson~\cite{Higgs,Djouadi:2005gi}.  A critical  question would be  whether such
a Higgs boson mass value allows to extrapolate the SM  up to ultimate scales,
while still having an absolutely  stable electroweak vacuum
\cite{stability,Ellis:2009tp,Degrassi:2012ry}. 
Indeed, it is well known that top quark quantum corrections tend to drive the
quartic Higgs coupling $\lambda$, which in the SM is related to the  Higgs mass
by the tree-level expression $\lambda=M_H^2/2v^2$ where $v\approx 246$ GeV is
the Higgs field vacuum expectation value, to negative values which render the
electroweak vacuum unstable. 

A very recent analysis, including the
state-of-the-art  quantum corrections at next-to-next-to-leading order (NNLO)
that are relevant in this context gives for the condition of absolute stability
of the electroweak vacuum, $\lambda (M_P) \geq 0$,  when  the SM is extrapolated
up to the Planck  scale $M_P$ \cite{Degrassi:2012ry}  

\begin{equation} 
\label{eq:MH-bound}
M_H \,\geq\, 
129.6
+ 1.8\! \times \! \left( \frac{m_t^{\rm pole} - 173.2~{\rm GeV}} {0.9~{\rm GeV}} 
\right) 
- 0.5 \! \times \! \left( \frac{ \alpha_s (M_Z) -0.1184}{0.0007}  \right) 
\pm 1.0 ~~{\rm GeV}\, .
\end{equation} 
This full NNLO calculation is based on three main ingredients that have been
calculated only very recently: the two-loop threshold  corrections to the
quartic coupling $\lambda$ at the weak scale, $\lambda(\mu)= M_H^2/2v^2+\Delta
\lambda(\mu)$ which involve the QCD and the Yukawa interactions
\cite{Degrassi:2012ry,Bezrukov:2012sa},  the three-loop leading contributions to the
renormalization group evolution of the coupling $\lambda$ as well as the top
quark Yukawa coupling and the Higgs mass anomalous dimension
\cite{Chetyrkin:2012rz}, and  the three-loop corrections to the beta functions of
the three SM gauge couplings taking into account Yukawa and Higgs self
couplings \cite{Mihaila:2012fm}. The uncertainty of $\Delta M_H =\pm 1.0$ GeV quoted in eq.~(\ref{eq:MH-bound})
reflects the theoretical uncertainty on the  Higgs mass bound  which, to
a good  approximation, corresponds to the difference between the results
obtained  when calculating the bound at next-to-leading order (NLO) and
NNLO\footnote{
Note, that the vacuum stability analysis 
of Ref.~\cite{Bezrukov:2012sa} based on the two-loop ${\cal O}(\alpha \alpha_s)$ threshold corrections
and the three-loop terms in the renormalization group equation~\cite{Chetyrkin:2012rz,Mihaila:2012fm}, 
arrives at a similar relation for $M_H$ as in eq.~(\ref{eq:MH-bound}).
The additional improvements of Ref.~\cite{Degrassi:2012ry} due to 
${\cal O}(\alpha_s^2)$ terms shift the bound on $M_H$ by a small amount, of
order 0.1~GeV.}.

The vacuum stability condition eq.~(\ref{eq:MH-bound}) critically depends on
three basic  inputs. 

A first parameter is the Higgs boson mass $M_H$ which, from the current excess
of data,  seems to be in the (wide) range of $M_H \approx 124$-128 GeV
\cite{LHC}. 

A second one is the strong coupling  constant $\alpha_s$ evaluated at the scale
of the $Z$ boson mass, with a world average value of \cite{Nakamura:2010zzi}
\begin{equation} 
\label{eq:as-value}
\alpha_s(M_Z)=0.1184 \pm 0.0007
\, .
\end{equation}
The combined theoretical and experimental uncertainty of  $\Delta \alpha_s\! =
\! \pm 0.0007$  generates, at the 2$\sigma$ level, an uncertainty of $\Delta M_H
\approx 1$ GeV on the Higgs mass bound\footnote{ This value of $\alpha_s$ is
obtained from a large set of measurements with significant spreads between them.
This issue will be discussed later.}.

The most critical ingredient in eq.~(\ref{eq:MH-bound}) is the top quark pole
mass, identified with the one measured at the Tevatron by the CDF and D0
collaborations\footnote{In contrast to Ref.~\cite{Degrassi:2012ry}, we do not
average the $m_t$ value determined at the Tevatron with that obtained at the LHC
by the ATLAS and CMS collaborations which  presently have much larger
uncertainties \cite{Blyweert:2012bq}.  } \cite{Lancaster:2011wr},    
\begin{equation} 
\label{eq:mt-value}
m_t^{\rm exp}=173.2 \pm 0.9~{\rm GeV}  
\, .
\end{equation}
Indeed, a change of the input $m_t$ value by 1 GeV will lead to a $\Delta M_H
\approx \pm 2$~GeV variation of the Higgs mass bound.   Allowing for a 2$\sigma$
variation of the top quark mass value alone, one obtains the upper bound $M_H \!
\geq \! 125.6$ GeV.  Hence, if the Higgs mass were exactly $M_H=125$ GeV, the 
absolute stability of the electroweak vacuum up to the Planck scale would be
excluded at the 95\%  confidence level, while the value $M_H = 126$ GeV would
instead allow for the stability of the vacuum at the same confidence level.

Thus, the ``fate of the universe" \cite{Ellis:2009tp}, i.e. whether the
electroweak vacuum  is stable or not up to the largest possible high-energy
scale, critically relies on a precise determination of the Higgs boson and top
quark masses (besides the strong coupling constant). While it is expected  that
the clean $H \to \gamma \gamma$ and $H \to ZZ \to 4\ell^\pm$ (with $\ell=e,\mu)$
decay  channels and the excellent energy resolution for photons and charged
leptons should allow to ultimately measure the Higgs boson  mass with a
precision of ${\cal O}(100)~{\rm MeV}$ \cite{TDRs}, severe
theoretical and experimental problems occur in the case of the top quark
mass\footnote{ This situation is similar to that occurring in the context of 
the electroweak precision tests and the indirect determination of the Higgs
mass, where the definition of the top  mass also plays a key role.  However,
while the impact of $m_t$ is relatively modest in  the global electroweak  fits
as the  resulting Higgs mass value has a large uncertainty, $\Delta M_H
\!\approx \!30$ GeV \cite{Bardin:1999yd}, it is extremely  strong for the
stability bound.}.  

An immediate problem is that the top quark mass parameter measured at the
Tevatron (and to be measured at the LHC) via kinematical  reconstruction from
the top quark  decay  products and comparison to Monte Carlo simulations,  is
not necessarily the pole mass which should enter the stability  bound eq.~(\ref{eq:MH-bound}).  
Besides the fact that  the reconstruction of the coloured top quark  four
momentum from its  uncolored decay products introduces an intrinsic uncertainty
due to the non-perturbative mechanism of hadronisation that can be hardly
quantified, there is an important conceptual problem. Strictly speaking, a
theoretical prediction of a given  measured  observable is required to extract a
parameter of a model in a meaningful way and this prediction should be made 
beyond the leading-order  approximation for which a renormalisation scheme can
be fixed. Obviously,  this is not the case for the mass currently measured at
the Tevatron which is merely the mass parameter in a Monte Carlo program  with
which the kinematical fit  of the top decay products is performed and which does
resort to any given  renormalisation scheme. 

 Furthermore, it is well known that the concept of an  ``on-shell" or ``pole"
quark mass has intrinsic theoretical limitations as quarks are  colored objects
and, as such, do not appear as asymptotic states of the S-matrix because of
color confinement \cite{ambiguity}.  In addition, because of the so-called
infrared renormalons, such a pole mass is plagued with  an intrinsic 
non-perturbative ambiguity of the order of $\Lambda_{\rm QCD}$ amounting to a few hundred
MeV, and it cannot be ``measured" with an accuracy better\footnote{A precise
quantitative statement is rather  hard to make and more work in this direction
is needed. Very few studies have been devoted to the relations between the
``Monte Carlo", the experimentally measured and the ``pole" quark mass.  In
Ref.~\cite{Hoang:2008xm}, the uncertainties due to non-perturbative  color
reconnection effects in the hadronisation process and from the ambiguities in
the top quark mass definition were estimated to be of order $\pm 0.5$ GeV each. 
In Ref.~\cite{Skands:2007zg}, the combined  effect of color-reconnection, underlying
events and shower in the Monte Carlo programs was estimated in a toy model to
generate an uncertainty of $\pm 1.5$ GeV on the reconstructed top quark mass at
the Tevatron. These effects are not included in the central value and error for
the top quark mass quoted in eq.~(\ref{eq:mt-value}).}  than ${\cal O}(\Lambda_{\rm QCD})$ 
\cite{ambiguity}.  

So-called short distance top quark masses, such as the one defined in the
modified minimal substraction ($\overline{\rm MS}$)  scheme at a scale $\mu$,
$m_t^{\overline{\rm MS}} (\mu)$, offer remedy to these problems. The
$\overline{\rm MS}$ mass realizes the concept of a  running mass which depends
on the hard scale $\mu$ of the process  in complete analogy to the running
coupling $\alpha_s(\mu)$.  A  determination of $m_t^{\overline{\rm MS}} (\mu)$ is
then possible from the mass dependence  of any observable which is rather
precisely measured and, at the same time, theoretically predicted beyond the
leading order (LO) approximation in QCD perturbation theory. An immediate choice
for the determination of $m_t^{\overline{\rm MS}} (\mu)$  is the total
production cross section for top quark pairs, $\sigma(t\bar t\!+\!X)$.  
It has been measured both at the Tevatron and the LHC with an accuracy of
better than 10\%  and it is known to very good approximation at NNLO in QCD  in
the convenient $\overline{\rm MS}$ renormalisation 
scheme~\cite{Moch:2008qy,
Langenfeld:2009wd,Moch:2012mk,Baernreuther:2012ws}.
The most recent combinations of inclusive cross section measurements at the
Tevatron performed  by the CDF and D0 collaborations yield a
value~\cite{cdf:2009note,Abazov:2011cq},
\begin{eqnarray} 
\label{eq:crstev}
\sigma( p\bar p \to t\bar t\! + \!X) \, = \, 
\renewcommand{\arraystretch}{1.2}
\begin{array}{l} 
7.56\,^{+0.63}_{-0.56}~{\rm pb~(D0) \ \ and \ \ } 
7.50\,^{+0.48}_{-0.48}~{\rm pb~(CDF)} 
\, .
\end{array} 
\end{eqnarray}
At the LHC in the run with a center-of-mass energy of $\sqrt s \!=\! 7$ TeV
the ATLAS and CMS collaborations have each measured a combined cross section~\cite{atlas:2011note,cms:2011note} 
of
\begin{eqnarray} 
\label{eq:crslhc7}
\sigma(pp \! \to \! t\bar t \!+\!X)\ =  
\renewcommand{\arraystretch}{1.2}
\begin{array}{l}
177^{+11}_{-10}~{\rm pb~(ATLAS) \ \ and \ \ } 
165.8^{+13.3}_{-13.3}~{\rm pb~(CMS)} \, .
\end{array} 
\end{eqnarray}

The first issue that we will address in the present paper is the  comparison of
the cross section measurements above  with the theory predictions, which will 
allow us to extract the $\overline{\rm MS}$ top quark mass $m_t^{\overline{\rm
MS}}$  and use it subsequently to derive the pole top quark mass and, hence, 
the vacuum stability bound eq.~(\ref{eq:MH-bound}) in an unambiguous way. To
that end we update the analyses of
Refs.~\cite{Langenfeld:2009wd,Abazov:2011pta}  using the latest sets of parton
distribution functions (PDFs) at 
NNLO~\cite{Alekhin:2012ig,JimenezDelgado:2009tv,Martin:2009iq,Ball:2011uy} and,
most importantly, the new NNLO QCD contributions to $\sigma( p\bar p \to t\bar t
+X)$  in the high-energy limit~\cite{Moch:2012mk}  and for the $q\bar q \to
t\bar t+X$ channel \cite{Baernreuther:2012ws}.  

In a second  part of this paper, we recall that a self-consistent and precise
determination  of the top quark mass can best be performed at a high-energy
electron-positron collider, especially when scanning the kinematical threshold
for $t\bar t$ pair production.  The accuracy that can be achieved on short
distance masses such as  the 1S-threshold mass amounts to $\Delta m_t^{\rm 1S}
\approx 100$ MeV~\cite{Hoang:2000yr,Hoang:2001mm}.  Together with a Higgs mass
measurement with a comparable accuracy  or less and a more precise determination
of the strong coupling $\alpha_s$  this would ultimately allow to verify the
stability bound in the SM  at the few per mille level.\bigskip 

Let us briefly summarize how to obtain the top quark pole mass $m_t^{\rm pole}$
from  the total production cross section $\sigma( p\bar p/pp \to t\bar t +X)$ at
hadron colliders.  This observable has been computed to very good approximation 
at NNLO in QCD based on the large threshold 
logarithms~\cite{Moch:2008qy,
Langenfeld:2009wd}
which provide sufficiently precise phenomenological predictions in the parton 
kinematic range covered by the Tevatron and the LHC  with a center-of-mass energy of $\sqrt
s \!=\! 7$ TeV.  Most recently, the exact NNLO result for contributions  to
the $q\bar q \to t\bar t+X$ channel  \cite{Baernreuther:2012ws} and the
constraints imposed by the high-energy factorization have been derived~\cite{Moch:2012mk}.  
This knowledge suffices to predict  $\sigma( p\bar p \!\to\! t\bar t \!+\!X)$ at
Tevatron in eq.~(\ref{eq:crstev}) and $\sigma( pp \! \to \! t\bar t \!+\!X)$ at
the LHC in eq.~(\ref{eq:crslhc7}) with a few percent accuracy\footnote{We do
not account here for the electroweak radiative corrections  at
NLO~\cite{Beenakker:1993yr
}.  For light Higgs
bosons with  $M_H \approx 126$~GeV, these are vanishingly small at the Tevatron
and give a negative contribution of ${\cal O}(2\%)$ at the LHC. Bound state
effects and the resummation of Coulomb type corrections have been shown to be
small at the Tevatron  as
well~\cite{Hagiwara:2008df
}.   Likewise, we do not
include the electroweak radiative corrections derived  in
Ref.~\cite{Jegerlehner:2003py} in the conversion  of the pole mass  $m_t^{\rm
pole}$ to the running mass $m_t^{\overline{\rm MS}} (\mu)$.}.

Conventionally, higher order computations in QCD employ the pole mass scheme for
heavy quarks.  It is straightforward though, to apply the well-known  conversion
relations~\cite{Gray:1990yh
} which are known
even beyond NNLO in QCD  to derive the total cross section as a function of  the
$\overline{\rm MS}$ mass~\cite{Langenfeld:2009wd,Aliev:2010zk}. As a benefit of
such a procedure, one arrives at theoretical predictions  for hard scattering
cross sections with better convergence properties  and greater perturbative
stability at higher orders  in the case of the $\overline{\rm MS}$  mass. 
We use  the cross section predictions  obtained with the program {\tt
HATHOR} (version 1.3)~\cite{Aliev:2010zk}  at NNLO accuracy with the latest
improvements of Refs.~\cite{Moch:2012mk,Baernreuther:2012ws}. These are 
combined with modern sets of PDFs,  ABM11~\cite{Alekhin:2012ig},
JR09~\cite{JimenezDelgado:2009tv}, MSTW08~\cite{Martin:2009iq}, and
NN21~\cite{Ball:2011uy}  and account for the full theoretical uncertainties,
i.e., the scale variation as well as the (combined) PDF and $\alpha_s$
uncertainty.  From eq.~(\ref{eq:crstev}),   we obtain the
values given in Table~\ref{tab:mt-values-tev}  when the CDF and D0 cross section
measurements are combined.  
\begin{table}[ht!] 
\begin{center} \renewcommand{\arraystretch}{1.5}
\begin{tabular} {|l|c|c|c|c|} \hline CDF\&D0  & ABM11 & JR09 & MSTW08 & NN21
\\ \hline  
${m_t^{\overline{\rm MS}}(m_t)}$
& 162.0$\,^{+2.3}_{-2.3}\,^{+0.7}_{-0.6}$  &
163.5$\,^{+2.2}_{-2.2}\,^{+0.6}_{-0.2}$   &
163.2$\,^{+2.2}_{-2.2}\,^{+0.7}_{-0.8}$   &
164.4$\,^{+2.2}_{-2.2}\,^{+0.8}_{-0.2}$   \\ ${m_t^{\rm pole}}$               &
171.7$\,^{+2.4}_{-2.4}\,^{+0.7}_{-0.6}$  &
173.3$\,^{+2.3}_{-2.3}\,^{+0.7}_{-0.2}$   &
173.4$\,^{+2.3}_{-2.3}\,^{+0.8}_{-0.8}$   &
174.9$\,^{+2.3}_{-2.3}\,^{+0.8}_{-0.3}$   \\ (${m_t^{\rm pole}}$)
&(169.9$\,^{+2.4}_{-2.4}\,^{+1.2}_{-1.6}$)
&(171.4$\,^{+2.3}_{-2.3}\,^{+1.2}_{-1.1}$) 
&(171.3$\,^{+2.3}_{-2.3}\,^{+1.4}_{-1.8}$) 
&(172.7$\,^{+2.3}_{-2.3}\,^{+1.4}_{-1.2}$)  \\ \hline 
\end{tabular} 
\end{center}
\vspace*{-3mm} 

\caption{ \label{tab:mt-values-tev} \small The value of the top
quark mass ${m_t^{\overline{\rm MS}} (m_t)}$ in GeV at NNLO in QCD  determined
with four sets of NNLO PDFs 
from the measurement of $\sigma( p\bar p \to t\bar t+X)$ at the Tevatron
when the CDF and D0 results quoted in eq.~(\ref{eq:crstev}) are combined. 
The set of uncertainties originate from the experimental error on $\sigma( p\bar p \to
t\bar t+X)$ (first error) and from the variation of the factorization and
renormalization scales  from $\frac12 m_t \leq \mu_F=\mu_R \leq 2m_t$ (second
error). The resulting pole mass ${m_t^{\rm pole}}$ in the second line is
obtained from a scheme transformation to NNLO accuracy, using the program {\tt
RunDec}  and the value of $\alpha_s(M_Z)$ of the given
PDF set. For comparison, in the third line in parentheses, $({m_t^{\rm pole}})$
is also given as  extracted directly from the measured cross section. }
\vspace*{-3mm} 
\end{table} 

The values for  $m_t^{\overline{\rm MS}}(m_t)$ in Table~\ref{tab:mt-values-tev} 
determined from the combined Tevatron cross sections 
carry an uncertainty of  $\Delta_{\rm exp} m_t^{\overline{\rm MS}}(m_t) \approx \pm
2.3$~GeV  due to the experimental errors in eq.~(\ref{eq:crstev}). The residual
scale dependence of the theory prediction for $\sigma( p\bar p \to t\bar t +
X)$, which is determined in the interval  $\frac12 m_t \! \leq \! \mu_F\! =\!
\mu_R \! \leq \! 2m_t$ as  effects due to $\mu_F \! \neq \! \mu_R$ are small at
NNLO \cite{Langenfeld:2009wd,Aliev:2010zk},  results in an error of
$\Delta_{\rm scale} m_t^{\overline{\rm MS}}(m_t) \! \approx \! \pm 0.7$~GeV  illustrating the great
stability of the perturbative expansion at NNLO in QCD  when using the running
$\overline{\rm MS}$ mass.

The second line in Table~\ref{tab:mt-values-tev} lists the pole mass values 
$m_t^{\rm pole}$ at NNLO obtained from the values for the $\overline{\rm MS}$
mass  $m_t^{\overline{\rm MS}}(m_t)$ using the scheme transformation given in
Ref.~\cite{Gray:1990yh
}  as implemented in the
program {\tt RunDec}~\cite{Chetyrkin:2000yt}  together with the $\alpha_s(M_Z)$
value of the given PDF set. For comparison, the third line in
Table~\ref{tab:mt-values-tev} quotes  in parentheses the value of $m_t^{\rm
pole}$ determined by a direct  extraction from the NNLO theory prediction using
the on-shell scheme. The differences of ${\cal O}(+2)$~GeV with the values in
the second line  obtained from converting the $\overline{\rm MS}$ mass  indicate
the importance of higher order corrections beyond NNLO in QCD  if using the pole
mass scheme. This is to be contrasted with the observed very good apparent
convergence  of the perturbative predictions already at NNLO  in the running
mass $m_t^{\overline{\rm MS}}(m_t)$ scheme,  see Ref.~\cite{Langenfeld:2009wd}.

There is one particular aspect in the chosen procedure, though, which requires attention.
The Tevatron cross section data~\cite{cdf:2009note,Abazov:2011cq} 
acquire a weak dependence on the top quark mass in the extrapolation from 
the recorded events in the fiducial volume to the total cross section.  
This is induced by comparison to Monte Carlo simulations 
and the values quoted in eq.~(\ref{eq:crstev}) assume a mass of $m_t = 172.5$~GeV. 
This systematic uncertainty of $\sigma( p\bar p \to t\bar t + X)$ 
has been published by the D0 collaboration~\cite{Abazov:2011cq} 
as a parametrization in $m_t$. 
For CDF, it has not been published for the value in eq.~(\ref{eq:crstev}) 
based on the combination of data at 4.6~fb$^{-1}$ luminosity in~\cite{cdf:2009note}. 
It has, however, been quoted as a shift for $\sigma( p\bar p \to t\bar t + X)$ 
of approximately $\Delta \sigma / \sigma \approx -0.01 \Delta m_t$/GeV 
in a previous combination of data at 760~pb$^{-1}$ luminosity~\cite{cdf:2006note}.
In order to account for this additional source of systematic uncertainty 
one can identify this parameter $m_t$ with the on-shell mass and check that 
the pole mass values in Table~\ref{tab:mt-values-tev} are consisted with $m_t = 172.5$~GeV 
within $\Delta_{\rm sys} m_t \approx \pm 1$~GeV.
This assumption is motivated by the fact, that the NLO computations applied in the
experimental analysis, e.g., {\tt MC@NLO}~\cite{Frixione:2003ei} 
or {\tt MCFM}~\cite{Campbell:2002tg
} contain 
perturbative matrix elements at NLO in QCD using the pole mass scheme for the top quark.
At the moment however, we are lacking further quantitative information. 
Therefore it is reassuring to see that the potential shifts of 
$\Delta_{\rm sys} m_t \approx \pm 1$~GeV are contained well within the experimental error on $m_t^{\rm pole}$
in Table~\ref{tab:mt-values-tev}.

The largest residual uncertainty in the extraction of $m_t^{\overline{\rm
MS}}(m_t)$  in Table~\ref{tab:mt-values-tev} resides in the dependence on the
PDFs as can be seen by comparing the central values for the sets ABM11, JR09, MSTW
and NN21. Although the $q \bar q$ parton luminosity is quite well  constrained
in the kinematical range of interest at the Tevatron,  the differences in the
individual global fits (value of $\alpha_s(M_Z)$ etc.)  lead to a spread in the
central value of $m_t^{\overline{\rm MS}}(m_t) \approx 162.0$ GeV to $164.4$~GeV. 
This is larger than the combined PDF and $\alpha_s$ uncertainty   of any
individual set not quoted in Table~\ref{tab:mt-values-tev} 
which amounts to an additional error of  $\Delta_{\rm PDF}
m_t^{\overline{\rm MS}}(m_t) \approx \pm 0.7$~GeV,  except for JR09, where one finds $\Delta_{\rm
PDF} m_t^{\overline{\rm MS}}(m_t) \approx \pm 1.4$~GeV. Yet, within the PDF
uncertainty the values of $m_t^{\overline{\rm MS}}(m_t)$  in
Table~\ref{tab:mt-values-tev} are largely consistent at the level of 1$\sigma$.

Combining the $m_t^{\overline{\rm MS}}(m_t)$ values in Table~\ref{tab:mt-values-tev} 
from the combined Tevatron measurements, 
we obtain the central value of the $\overline{\rm MS}$ mass 
and its associated uncertainty at NNLO
\begin{equation}
m_t^{ \overline{\rm MS}}(m_t) \,=\, 163.3 \pm 2.7~{\rm GeV}
\, ,
\end{equation}
which is equivalent to the top quark pole mass value of
\begin{equation}
m_t^{\rm pole} \,=\, 173.3 \pm 2.8~{\rm GeV}
\, ,
\end{equation}
where all errors were added in quadrature including the $\Delta_{\rm sys}  m_t
\!\approx \!\pm 1$~GeV discussed above. Note that, although the total error is 
a factor of four larger,   the central value is remarkably close to that of 
$m_t^{\rm exp}$ in eq.~(3) determined from the top decay products at
the Tevatron.  

When injected in eq.~(\ref{eq:MH-bound}), the value of the top pole mass
above\footnote{One could  write directly  eq.~(1) in terms of the $\overline{\rm
MS}$ top quark mass which is in fact the basic input entering the top Yukawa
coupling which is defined in the $\overline{\rm MS}$ scheme.  This will prevent
the unnecessary translation to the pole mass both in  the stability bound and in
the $p\bar p \to t\bar t$ cross section. Such a formula will be soon provided 
by the authors of Ref.~\cite{Degrassi:2012ry}. We thank Gino Isidori for a
discussion on this point.} leads to the upper bound for vacuum stability to be
realized (ignoring the theoretical and the experimental uncertainties on the
Higgs mass and on $\alpha_s$)  
\begin{equation}
M_H \,\geq\,129.8 \pm 5.6~{\rm GeV}
\, ,
\end{equation}
in which the Higgs mass values $M_H\approx 124$-127 GeV, indicated by the 
ATLAS and CMS searches, comply in contrast to the case where the 
mass value of eq.~(\ref{eq:mt-value}) from kinematical reconstruction 
at the Tevatron is used instead.  
Note also that the uncertainties are much larger, a factor of approximately $3$,  
than if eq.~(\ref{eq:mt-value}) were used. 

Let us now turn to the LHC.  The top quark mass extracted from the combined
ATLAS and CMS  measurements in eq.~(\ref{eq:crslhc7})   is given in
Table~\ref{tab:mt-values-lhc7}.  While the uncertainty 
$\Delta_{\rm exp} m_t^{\overline{\rm MS}}(m_t) \approx \pm 2.3$~GeV  due to the experimental
errors is similar to the Tevatron measurement the theoretical uncertainty due to
the scale variation is mostly larger, i.e., 
$\Delta_{\rm scale} m_t^{\overline{\rm MS}}(m_t) \approx \pm 1.1$~GeV. 
The most striking observation in
Table~\ref{tab:mt-values-lhc7}  is certainly the very large spread in the
central value of  $m_t^{\overline{\rm MS}}(m_t) \approx 159.0$ GeV to
$166.7$~GeV depending on the chosen PDF set. The combined PDF and $\alpha_s$
uncertainty of the individual sets in Table~\ref{tab:mt-values-lhc7}  is in the
range $\Delta_{\rm PDF} m_t^{\overline{\rm MS}}(m_t) \approx \pm 1.0$~GeV to  
$1.4$~GeV and  $\Delta_{\rm PDF} m_t^{\overline{\rm MS}}(m_t) \approx \pm
2.4$~GeV for JR09. This leads to consistency between the central values for
$m_t^{\overline{\rm MS}}(m_t)$  for each PDF set (comparing Tevatron in 
Table~\ref{tab:mt-values-tev} and LHC in Table~\ref{tab:mt-values-lhc7}) 
but the ones obtained for the different PDF sets at the LHC 
are not compatible with each other within the errors.

\begin{table}[ht!] 
\small \begin{center}
\renewcommand{\arraystretch}{1.4} \begin{tabular} {|l|c|c|c|c|} \hline 
ATLAS\&CMS 
& ABM11 & JR09 & MSTW08 & NN21 
\\ \hline  ${m_t^{\overline{\rm MS}}(m_t)}$ &
159.0$\,^{+2.1}_{-2.0}\,^{+0.7}_{-1.4}$  &
165.3$\,^{+2.3}_{-2.2}\,^{+0.6}_{-1.2}$   &
166.0$\,^{+2.3}_{-2.2}\,^{+0.7}_{-1.5}$   &
166.7$\,^{+2.3}_{-2.2}\,^{+0.8}_{-1.3}$   \\ ${m_t^{\rm pole}}$               &
168.6$\,^{+2.3}_{-2.2}\,^{+0.7}_{-1.5}$  &
175.1$\,^{+2.4}_{-2.3}\,^{+0.6}_{-1.3}$   &
176.4$\,^{+2.4}_{-2.3}\,^{+0.8}_{-1.6}$   &
177.4$\,^{+2.4}_{-2.3}\,^{+0.8}_{-1.4}$   \\ (${m_t^{\rm pole}}$)
&(166.1$\,^{+2.2}_{-2.1}\,^{+1.7}_{-2.3}$)
&(172.6$\,^{+2.4}_{-2.3}\,^{+1.6}_{-2.1}$) 
&(173.5$\,^{+2.4}_{-2.3}\,^{+1.8}_{-2.5}$) 
&(174.5$\,^{+2.4}_{-2.3}\,^{+2.0}_{-2.3}$)  \\ \hline 
\end{tabular} \end{center}
\vspace*{-3mm} 
\caption{ \label{tab:mt-values-lhc7} \small Same as
Table~\ref{tab:mt-values-tev} for  the measurement of $\sigma( p p\! \to 
\! t\bar t\! + \! X)$ at the LHC with $\sqrt s=7$ TeV when the ATLAS
and CMS results in eq.~(\ref{eq:crslhc7}) are combined. } 
\vspace*{-1mm} 
\end{table}

Also the LHC experiments assume in eq.~(\ref{eq:crslhc7}) a mass of $m_t =
172.5$~GeV  when extrapolating the number of measured events with top quark
pairs  to the inclusive cross section $\sigma(pp \to t\bar t + X)$. However, no
information on the $m_t$ dependence of this procedure is given  in
Refs.~\cite{atlas:2011note,cms:2011note} and the same self-consistency  check
applied above by comparing to the pole mass value $m_t^{\rm pole}$  in the
second line of Table~\ref{tab:mt-values-lhc7}  shows that one should expect
a significantly larger systematic uncertainty. Thus, at present the determination
of $m_t^{\overline{\rm MS}}(m_t)$ from the inclusive cross section at LHC is
very difficult,  predominantly because of lacking information on the
experimental systematics and because of the strong correlation of the top quark
mass with the  value of $\alpha_s$ and the $gg$ parton luminosity in the theory
predictions. The latter problem could be addressed by combining measurements of
different  observables, for instance, by using a novel method for the top quark mass
determination from the ratio of rates for the process $t \bar t+$jets~\cite{Alioli:2012hj}.

The ultimate precision on the top quark mass to be reached at the LHC, however,
is hard to predict at the moment.  A total uncertainty  that is a factor of two
smaller than the present uncertainty from  the Tevatron measurements, 
\begin{equation}
\Delta m_t^{\rm pole}|_{\rm LHC\!-\!expected} \approx 1.5~{\rm GeV} 
\, ,
\end{equation}
does not seem to be excluded at present, but more work is  needed
to reach this level.

A very precise and unambiguous determination of the top quark mass and, hence, 
the possibility to derive a reliable upper bound on the Higgs mass for which 
the electroweak vacuum would be stable, can only be performed at an $e^+e^-$ 
collider ILC with an energy above $\sqrt
s\!=\!350$~GeV~\cite{Accomando:1997wt
}. 
Indeed, as a consequence of its large total decay width, $\Gamma_t\sim 1.5$~GeV, the top quark will decay before it hadronises
making  non-perturbative effects rather small and allowing to calculate quite
reliably the energy dependence of the  $e^+e^- \to t\bar t $ production cross
section when an energy scan is performed near the $t\bar t$ kinematical
threshold. The location of the cross  section rise allows to extract the value
of the  1S-threshold top quark mass, while the shape and normalization provide
information on the total width $\Gamma_t$ and on the strong coupling  $\alpha_s$
\cite{Martinez:2002st}.

The cross section $\sigma(e^+e^- \to t\bar t)$ at threshold is  known up to the
next-to-next-to-leading-logarithm (NNLL) using renormalization group
improvements and the next-to-next-to-next-to-leading order(N$^3$LO)  in the QCD
coupling is almost complete~\cite{Hoang:2000yr,Hoang:2001mm}.  It could
ultimately be determined  with a theoretical uncertainty of $\Delta \sigma (e^+
e^- \to t\bar t) \approx 3\%$ (the experimental uncertainties are much smaller)
but, as  the impact on the threshold top quark mass determination is rather
modest, an accuracy on $m_t$  much below 100 MeV can be achieved.   This
threshold mass can then be translated into the $\overline{\rm MS}$ top quark
mass $m_t^{\overline{\rm MS}}$(which  can be directly used  as input in the
Higgs mass stability  bound equivalent to eq.~(\ref{eq:MH-bound}) but in the
$\overline{\rm MS}$  scheme) or the one in the on-shell scheme, $m_t^{\rm
pole}$.   The combined experimental and theoretical uncertainty on the mass
parameter $m_t^{\rm pole}$ at the ILC  determined in this way, i.e., by
conversion form a short-distance mass,  is estimated to be 
\cite{Hoang:2000yr,Hoang:2001mm}
\begin{eqnarray}
\label{eq:mt-ilc}
m_t^{\rm pole}|_{\rm ILC} \lsim 200~{\rm MeV}
\, ,
\end{eqnarray}
i.e., an order of magnitude better than what can be achieved at the Tevatron and
the LHC.   In other words, the uncertainty in the top quark mass determination
will be so small at the ILC that its impact on the stability bound
eq.~(\ref{eq:MH-bound}) will become very mild. At such level of accuracy, the 
two parameters which will then be of concern are $M_H$ and $\alpha_s$. 

At a high-energy $e^+ e^-$ collider, the Higgs mass can be measured with an
accuracy below 100 MeV, and most probably $\Delta M_H \approx 50$ MeV, from the
recoil of the $Z$ boson  in the  Higgs-strahlung process $e^+ e^- \to HZ \to H
\ell^+\ell^-$ independently of the Higgs decays
\cite{Accomando:1997wt
}.

At the $e^+e^-$ collider, $\alpha_s$  can be  determined with an
accuracy close to or better than the one  currently adopted (which cannot be
considered to be  conservative\footnote{The world average $\alpha_s(M_Z)$ value
quoted in eq.~(\ref{eq:as-value})  is based on a comparison of QCD theory
predictions at least to NNLO accuracy with data on a variety of measurements
including  jet rates and event shapes in $e^+e^-$-collisions,  deep-inelastic
scattering (DIS), $Z$- and $\tau$-decays as well as entirely non-perturbative 
predictions based on  lattice simulations. The very small uncertainty of $\Delta
\alpha_s\! = \! \pm 0.0007$  is remarkable as recent high precision
determinations of $\alpha_s(M_Z)$  have lead to results which are only
marginally compatible within their quoted errors. This is the case for
$\alpha_s$ extractions from
$e^+e^-$-annihilation, see e.g.,~\cite{Gehrmann:2009eh
} or those based on DIS data~\cite{Alekhin:2009ni
}.
These differences can arise from theory assumptions such as  power corrections,
hadronisation corrections and so on and, likewise,  on the treatment of data,
see e.g., Ref.~\cite{Alekhin:2012ig} for a comparative study in the case of DIS.
In Ref.~\cite{pdg:2012} they have simply been averaged in an arithmetic manner.
Therefore, the uncertainty due to $\alpha_s$ attached to $M_H$ in
eq.~(\ref{eq:MH-bound}), should be considered at present as a lower bound at
most. If instead, one adopts the value $\alpha_s (M_Z)=0.1189 \pm 0.0026$  of
Ref.~\cite{Baikov:2012er} that has been determined  from $Z\to
q\bar q$ data and predicted to N$^{3}$LO accuracy in QCD 
(and which can be considered to be safe from short-comings of other analyses) 
one would have an uncertainty that is $\approx 4$ times  larger than in the case of the world
average eq.~(\ref{eq:as-value}), generating an uncertainty $\Delta M_H \approx
2$~GeV  on the Higgs mass bound eq.~(\ref{eq:MH-bound}) at the 1$\sigma$
level.})  $\Delta \alpha_s=0.0007$ \cite{Nakamura:2010zzi}, in a single
measurement; a statistical  accuracy of $\Delta \alpha_s=0.0004$ is for instance
quoted in Ref.~\cite{ILC-winter}. This can be done either in $e^+ e^- \to q\bar
q$ events  on the $Z$-resonance (the so-called GigaZ option) or at high energies
\cite{Accomando:1997wt
}
or in a combined fit with the top quark mass and total width in a  scan around
the $t\bar t$ threshold~\cite{Martinez:2002st}. 

Assuming for instance that accuracies of about $\Delta m_t \approx 200$ MeV and
$\Delta \alpha_s \approx 0.0004$ can be achieved at the ILC, a (quadratically)
combined uncertainty of less than $\Delta M_H \approx 0.5$ GeV on the Higgs mass
bound eq.~(\ref{eq:MH-bound}) could be reached. This would be of the same order
as  the  experimental uncertainty, $\Delta M_H \lsim 100$ MeV, that is expected
on the Higgs mass. 

At this stage we will be then mostly limited by the theoretical uncertainty in
the determination of  the stability bound eq.~(\ref{eq:MH-bound}) which is about  $\pm 1$ GeV.
The major part of this uncertainty originates from the the QCD threshold
corrections to the coupling $\lambda$ which are  known at the two-loop
accuracy \cite{Degrassi:2012ry,Bezrukov:2012sa}. It is conceivable that, by the time
the ILC will be operating, the theoretical uncertainty will decrease provided more
refined calculations of these threshold corrections  beyond NNLO are performed.  

The situation is illustrated in Fig.~1 where the areas for absolute stability, 
metastability\footnote{This situation occurs when  the true minimum of the
scalar potential is deeper than the standard electroweak minimum but the latter
has a lifetime that is larger than the age of the universe \cite{Ellis:2009tp}. The boundary for
this region is also taken from Ref.~\cite{Degrassi:2012ry}.} and instability of
the electroweak vacuum are displayed  in the $[M_H,m_t^{\rm pole}]$ plane at the
95\% confidence level. The boundaries are taken from 
Ref.~\cite{Degrassi:2012ry} but we do not include additional lines to account
for the theoretical uncertainty  of $\Delta M_H=\pm 1$ GeV (which could be
reduced in the future) and ignore for simplicity the additional error from the 
$\alpha_s$ coupling. 

\begin{figure}[htpb]
\vspace*{-.2cm}
\begin{center}
\epsfig{file=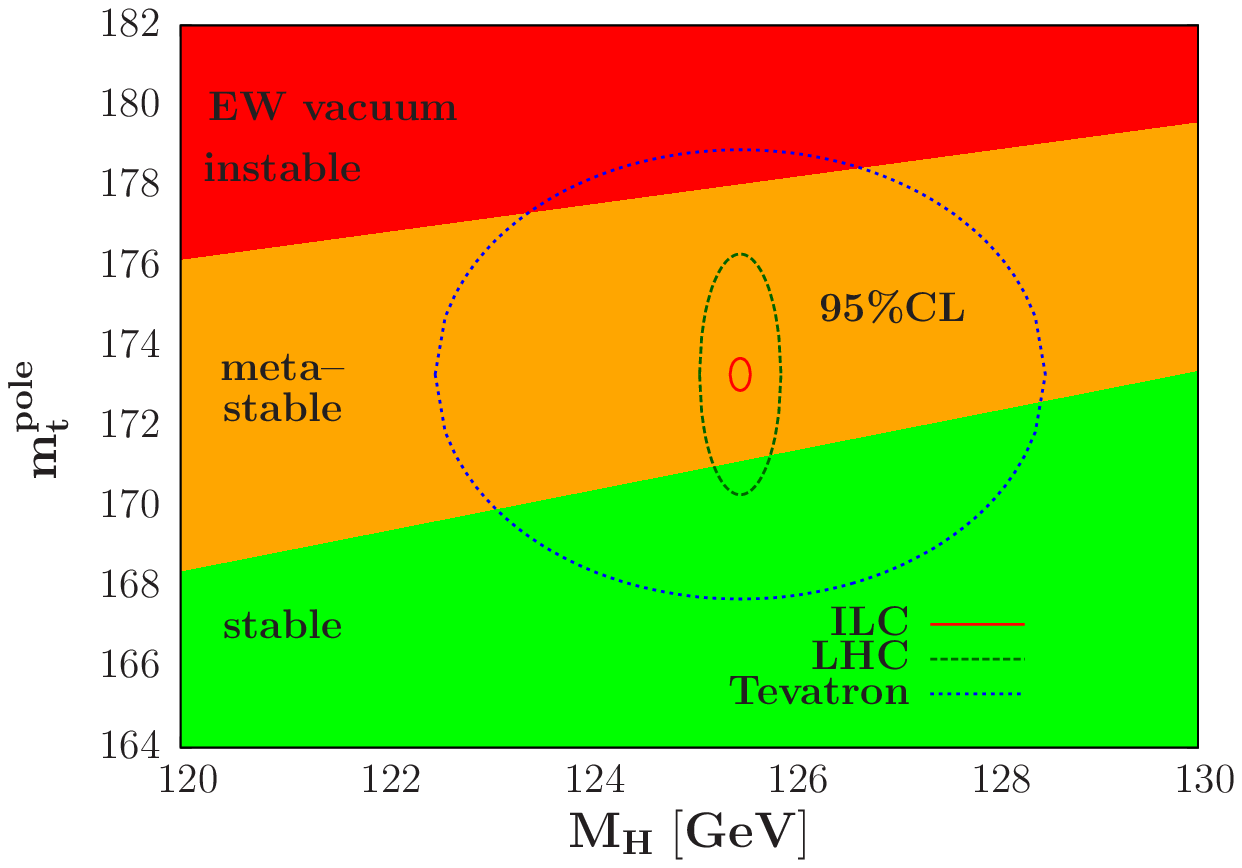,width=12.cm} 
\end{center}
\vspace*{-.7cm} 
\caption{\small The $2\sigma$ ellipses in the [$M_H, m_t^{\rm pole}$] plane that one
obtains from the current top quark and Higgs mass measurements at the Tevatron 
and LHC and which can be expected  in future measurements at the LHC and at
the ILC, when confronted with the areas in which the SM 
vacuum is absolutely stable, metastable and unstable up to the Planck scale. }
\vspace*{-2mm}
\end{figure}

As can be seen,  the $2\sigma$ blue--dashed ellipse for the present situation 
with the current Higgs and top quark masses of $M_H=126\pm 2$ GeV and 
$m_t^{\rm  pole}=173.3\pm 2.8$ GeV, and in which the errors are added in
quadrature, is large enough to cover the  three possibilities of absolute
stability, metastability and also instability. Assuming the same central values
as above, the green--dashed contour shows the impact  of an improved accuracy
on the top quark and Higgs masses of  $\Delta m_t^{\rm  pole}=\pm 1.5$ GeV and
$\Delta M_H = \pm 100$ MeV which is expected to be achieved at the LHC with more
accumulated data. With the present central values (which might of course change
with  more accurate measurements),  only the metastability and a small area of
the stability regions would be  covered.   The red--solid contour   represents
the expected situation at the ILC  where one could reach  accuracies of the
order of $\Delta M_H=\pm 50$ MeV on the Higgs mass and $\Delta m_t^{\rm 
pole}=\pm 200$ MeV on the top quark mass,  if obtained from a short-distance
mass, cf., eq.~(\ref{eq:mt-ilc}).  In this case, only one region, the
metastability region with the above assumed  central values, is covered (even
when the theoretical uncertainty on the bound is included). 
\bigskip

In conclusion, the present values of the Higgs boson mass as measured at the LHC
and the top quark pole mass as determined through a measurement of the  cross
section for top-quark pair production at the Tevatron, and that we have 
calculated in this paper to be $m_t^{\rm  pole}=173.3\pm 2.8$ GeV, are affected
with too large uncertainties which do not allow to draw a firm conclusion  on
the important question whether the electroweak vacuum is indeed stable  or not
when the Standard Model is extrapolated up to the Planck scale.  The situation
will not dramatically improve with a more accurate  measurement of the Higgs
boson  mass at the LHC as the top quark mass,  which plays the dominant role in
this issue, is not expected to be measured to better than $\pm 1.5$ GeV accuracy
even after a significant amount of LHC data. In particular, if the central 
$m_t^{\rm pole}$ value slightly moves downwards, it will be still undecided if
we are in the stable or metastable region.   It is only at a linear $e^+e^-$
collider where one could determine in a theoretically  unambiguous and 
experimentally very  precise way the top quark mass in a scan near the $e^+e^-
\to t\bar t$ kinematical threshold, and eventually  measure also more
accurately  the Higgs boson mass and the strong coupling constant  $\alpha_s$,
that the ``fate of the universe" could be ultimately decided. 
The importance of a future ILC in this respect has also been stressed 
in~\cite{Bezrukov:2012sa}.

If the measured central top quark and Higgs boson mass values turn out to be
such that one is close to the critical boundary for vacuum stability, which
implies that the Higgs self--coupling $\lambda$ and its $\beta_\lambda$ function
are very close to zero at the Planck scale, it would open a wide range of
interesting possibilities for new physics model building such as  
asymptotically safe gravitational theories \cite{Shaposhnikov1} or inflation
models that use the standard Higgs particle as the inflaton
\cite{Shaposhnikov2}. It is therefore very important that the intriguing
possibility $\lambda(M_P) \approx 0$ for the Higgs self-coupling is verified
experimentally in the most accurate manner. This provides a very strong argument
in favor of the most unambiguous and accurate  determination of the top quark
mass which plays a crucial role in this issue.\bigskip

\subsection*{Addendum:}
\setcounter{equation}{0}
\renewcommand{\theequation}{A.\arabic{equation}}

Several more precise measurements of the Higgs boson and top quark masses 
have been performed in the last few months, making an update of our 
analysis worthwhile.  

There is a first an update of $M_H$ as measured by ATLAS \cite{ATLAS-mH} and 
CMS \cite{CMS-mH} using the full $\approx 25$ fb$^{-1}$ data collected at 
$\sqrt s\!=\!7\!+\!8$ TeV and which leads to an average value
\begin{equation}
M_H \!  \simeq \! 125.6 \pm 0.4~{\rm GeV}
\end{equation}

In addition, an update of the top quark mass value as measured at the Tevatron with
up to 8.7 fb$^{-1}$ data has been recently released by the Tevatron electroweak working 
group.  Combining the statistical and systematic uncertainties leads to the 
value   \cite{Tevatron-mt}
\begin{equation}
m_{t}^{\rm exp} = 173.20 \pm 0.87~{\rm GeV}
\end{equation}

Finally, there was an recent reanalysis of the Tevatron and LHC measurements of the 
top--quark pair production cross sections from which an updated value of the top quark 
pole mass can be extracted.  When the experimental (statistical and systematic) error 
of $\pm 2.4$ GeV and the theoretical uncertainty of $\pm 0.7$ GeV from scale and PDF 
(keeping all correlations with the PDFs and the value of $\alpha_s$)  are added linearly,
one obtains \cite{new-mtpole}
\begin{equation}
m_{t}^{\rm pole} = 171.20 \pm 3.1~{\rm GeV}
\end{equation} 

Using these new inputs and fixing the strong coupling constant to $\alpha_s=0.1187$, the resulting contours in the [$M_H, m_t^{\rm pole}$] plane are confronted in Fig.~2 with the 
areas in which the SM vacuum is absolutely stable, meta-stable and unstable up to the 
Planck scale. 

\begin{figure}[htpb]
\vspace*{-3mm}
\mbox{\hspace*{-5mm}
\epsfig{file=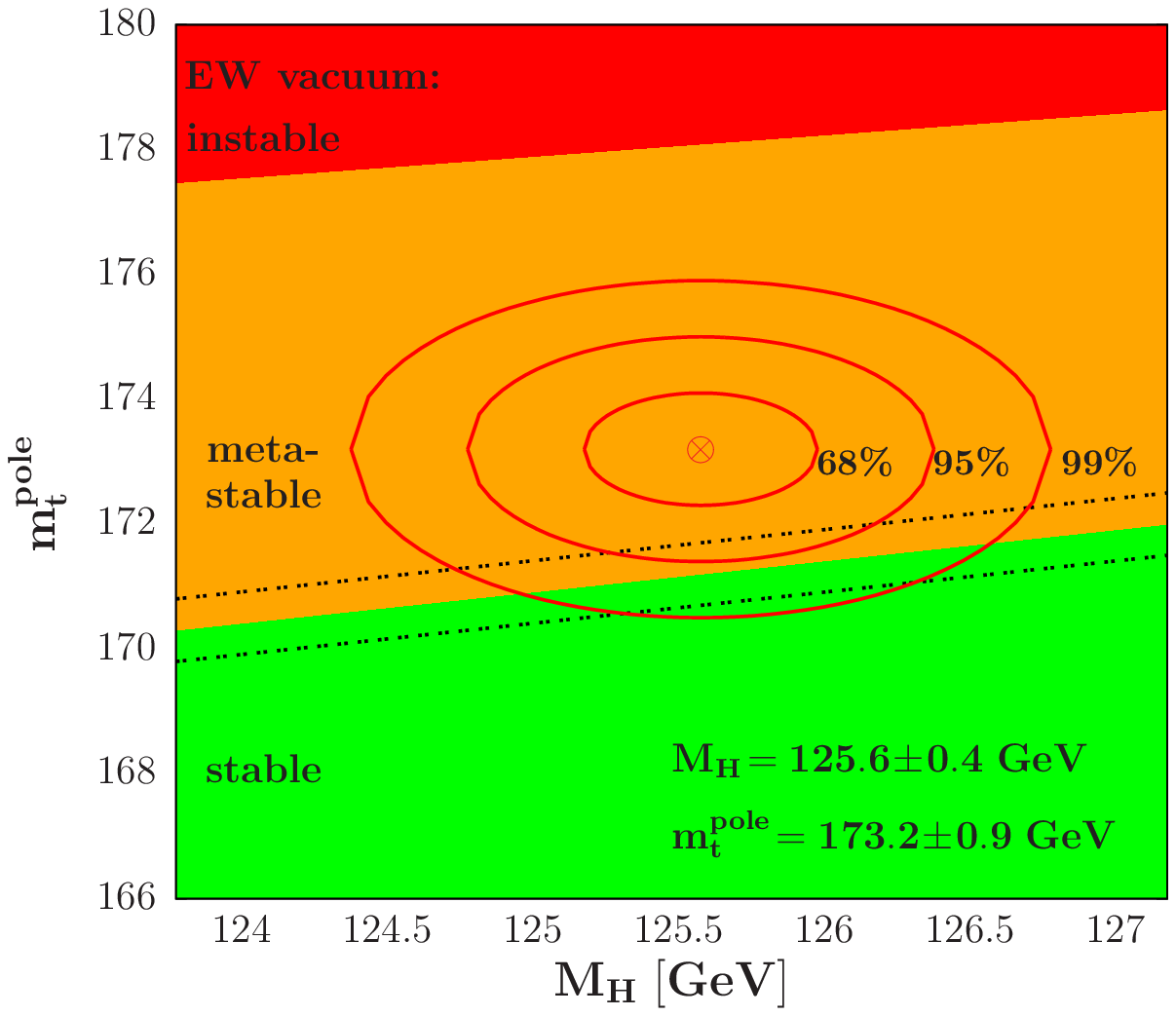,width=8.2cm} \hspace*{-5mm}
\epsfig{file=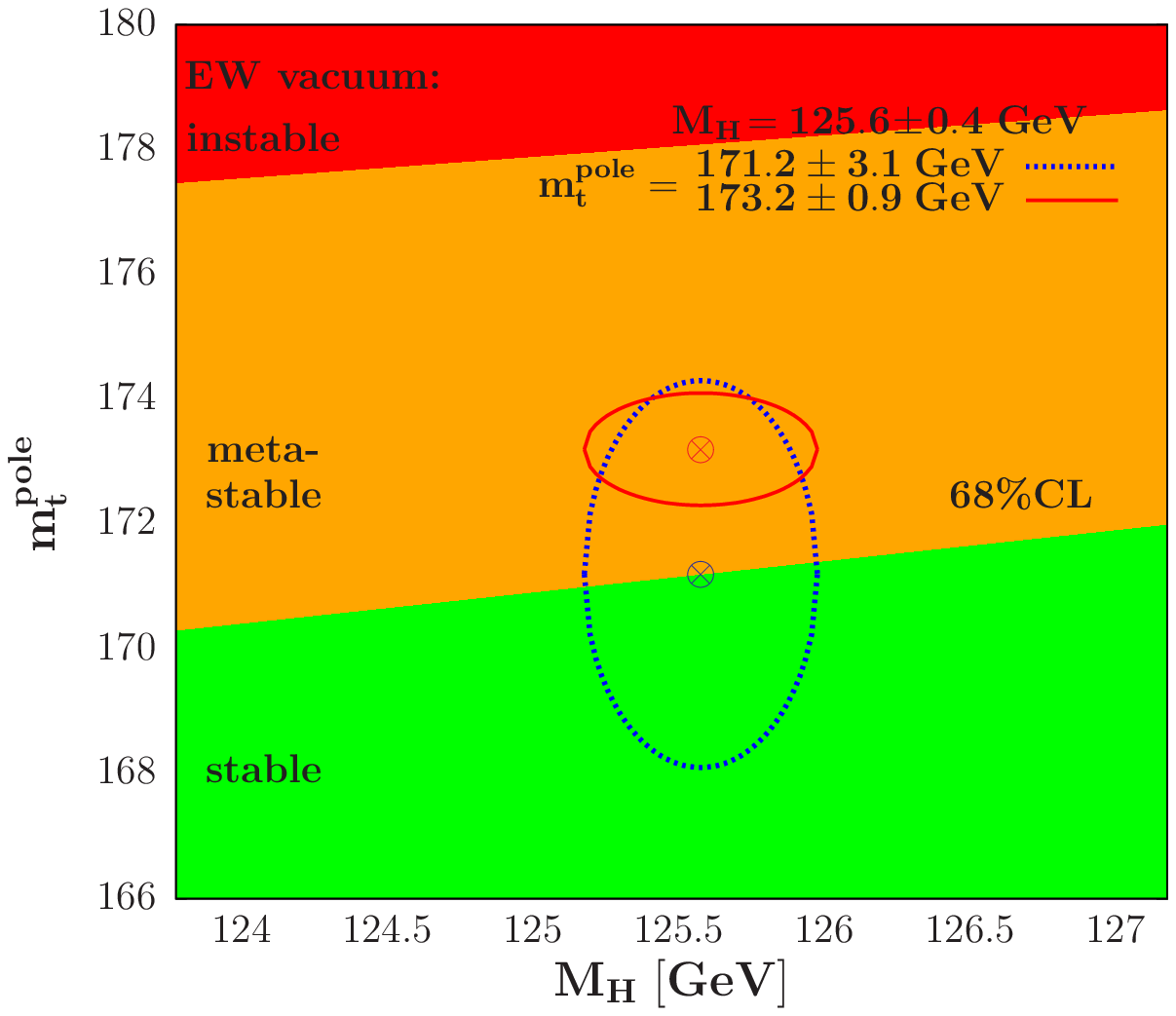,width=8.2cm} 
}
\vspace*{-.3cm} 
\caption{\small 
The ellipses in the [$M_H, m_t^{\rm pole}$] plane with the inputs $M_H=125.6 \pm 0.4$ GeV 
and $\alpha_s=0.1187$ are confronted with the areas in which the SM vacuum is absolutely 
stable, meta-stable and unstable up to the Planck scale. Left: the $1\sigma, 2\sigma$ 
and $3\sigma$ ellipses if $m_t^{\rm pole}$ is identified with the mass $m_t=173.2 \pm 
0.9$ GeV currently measured at the Tevatron; the black dotted lines indicate the theoretical 
uncertainty of $\Delta M_H=\pm 1$ GeV in the determination of the stability bound. Right: 
the $1\sigma$ ellipses when $m_t^{\rm pole}$ is identified with the one measured at the 
Tevatron and with the mass $m_t=171.2 \pm 3.1$ GeV extracted for the $t\bar t$ production 
cross section.}
\vspace*{-2mm}
\end{figure}

In the left-hand side of the figure, displayed are the $68\%, 95\%$ and $99\%$ 
confidence level contours if $m_t^{\rm pole}$ is identified with the mass measured 
at the Tevatron, $m_t=173.2 \pm 0.9$ GeV. It shows that at the $2\sigma$ level, the 
electroweak vacuum could be absolutely stable as the ellipse almost touches the
green area; this is particularly true if one includes the estimated theoretical uncertainty 
of $\Delta M_H=\pm 1$ GeV in the determination of the stability bound and  
indicated by the two black dotted lines in the figure. 

In the right--hand side of Fig.~2,  shown are the $68\%$CL contours when $m_t^{\rm 
pole}$ is identified with the one measured at the Tevatron and with the mass $m_t=171.2 
\pm 3.1$ GeV extracted for the $t\bar t$ production cross section. In the latter case, 
one sees that the central value of the top mass lies almost exactly on the boundary between 
vacuum stability and meta--stability. The uncertainty on the top quark mass is
nevertheless presently too large to clearly discriminate between these two possibilities.
\bigskip

\noindent {\bf Acknowledgments:} Discussions  with Gian Giudice and Gino Isidori
on Ref.~\cite{Degrassi:2012ry} and this manuscript   are gratefully
acknowledged. We also thank  J\'er\'emie Quevillon for helping to draw the
figure.$\;$AD thanks the CERN TH Unit for hospitality and support. This work has
been supported in part by the Deutsche Forschungsgemeinschaft in
Sonderforschungs\-be\-reich/Transregio~9,  by the European Commission through
contract PITN-GA-2010-264564 ({\it LHCPhenoNet}) and by the French ANR contract
TAPDMS {\it ANR-09-JCJC-0146}.

\end{document}